\documentclass[reprint,amsmath,amssymb,aps, twocolumn, superscriptaddress, prb]{revtex4-2}

\usepackage{hyperref}
\usepackage{comment}
\hypersetup{colorlinks,urlcolor=blue}
\usepackage{graphicx}
\usepackage[percent]{overpic}
\usepackage{dcolumn}
\usepackage{bm}
\usepackage{ulem}
\usepackage{physics}
\usepackage{lipsum}
\usepackage{appendix}
\usepackage{overpic}
\usepackage[T1]{fontenc}
\usepackage{tikz-network}

\definecolor{LB}{RGB}{134,41,198}

\tikzset{
  meter/.append style={draw, scale=0.6,inner sep=10, rectangle, font=\vphantom{A}, minimum width=30, line width=.8, 
    path picture={
      \draw[black] ([shift={(.1,.3)}] path picture bounding box.south west) 
        to[bend left=50] ([shift={(-.1,.3)}]path picture bounding box.south east);
      \draw[black,-latex] ([shift={(0,.1)}]path picture bounding box.south) -- 
        ([shift={(.3,-.1)}] path picture bounding box.north);
      }
    }
  }

\begin{document}
\preprint{APS/123-QED}
\title{Efficient tensor-network simulations of weakly-measured quantum circuits}
\author{Darren Pereira}
\affiliation{Laboratory of Atomic and Solid State Physics, Cornell University, Ithaca, New York 14853, USA}
\author{Leonardo Banchi}
\affiliation{INFN Sezione di Firenze, via G. Sansone 1, Sesto Fiorentino (FI) I-50019, Italy}
\affiliation{Department of Physics and Astronomy, University of Florence,
via G. Sansone 1, Sesto Fiorentino (FI) I-50019, Italy}

\begin{abstract}
  We present a tensor-network-based method for simulating a weakly-measured
  quantum circuit. In particular, we use a Markov chain to efficiently sample measurements and contract
  the tensor network, propagating their effect forward along the spatial
  direction. Applications of our algorithm include validating 
  quantum computers (capable of mid-circuit measurements) in
  regimes of easy classical simulability, and studying
  generative-machine-learning applications, where sampling from complex
  stochastic processes is the main task. 
  As a demonstration of our algorithm, we consider a (1+1)-dimensional
  brickwall circuit of Haar-random unitaries, interspersed with generalized
  single-qubit measurements of variable strength. We simulate the dynamics for
  tens to hundreds of qubits if the circuit exhibits area-law entanglement (under
  strong measurements), and tens of qubits if it exhibits volume-law
  entanglement (under weak measurements). We observe signatures of a
  measurement-induced phase transition between the two regimes as a function of
  measurement strength. 
\end{abstract}

\date{\today}
\maketitle

\section{Introduction}
\label{sec:Intro}

It has recently become possible to perform mid-circuit measurements on quantum-computing hardware \cite{chertkov2023characterizing,noel2022measurement,koh2023measurement,pan2020weak,agrawal2024observing,FengArxiv2025, GoogleMIPT2023}.
Although quantum algorithms can be dilated to move and perform all such measurements at the end of a circuit \cite{nielsen2010quantum}, 
this requires many ancillary qubits. By contrast, the measured qubits can be reused in mid-circuit measurements, 
allowing for a more efficient use of hardware. 
Aside from being an important milestone for implementing quantum error
correction \cite{devitt2013quantum} and mitigation \cite{botelho2022error}
strategies, mid-circuit measurements are a critical ingredient in a variety of
applications. In this work, we develop a classical algorithm to efficiently simulate and sample from variable-strength mid-circuit measurements.
We showcase our algorithm using one such application, the study of measurement-induced phase transitions (MIPTs) \cite{skinner2019measurement,JianPRB2020,BaoPRB2020, RQCAnnuRev2023,PotterVasseurChapter2022}.

MIPTs occur in systems which feature both entanglement generation and disruption, such as (partially) monitored quantum circuits. In a monitored quantum circuit, many-body unitary gates can generate a high degree of entanglement among qubits, but must compete with measurements, which destroy quantum correlations. Depending on the dominant effect, the result of these non-unitary dynamics is volume-law or area-law entanglement scaling of the system. MIPTs have been theorized to occur as a function of measurement rate \cite{skinner2019measurement, JianPRB2020,BaoPRB2020, RQCAnnuRev2023,PotterVasseurChapter2022} as well as measurement strength \cite{SzyniszewskiPRB2019,lee2023quantum, SuPRL2024,BulchandaniJStatPhys2024,AzizPRB2024,BaoPRB2020}. We use our algorithm to detect the presence of a MIPT as a function of measurement strength.

Our algorithm is written in the language of tensor networks. Tensor-network methods, such as those based on matrix-product states \cite{SchollwockAnnPhys2011},
enable the efficient simulation of quantum states with area-law entanglement in one dimension,
and have been applied to the study of MIPTs
\cite{TangPhysRevRes2020,DoggenPhysRevRes2022,CecilePhysRevRes2024,LoioPRB2025,BarrattPRL2022,MelkoPRA2021,YanayPRL2024}. Our approach improves on such works by
defining iterative circuit contractions that 
express the sampling problem as a Markov chain. This enables the efficient sampling  of measurement outcomes 
using $\mathcal O(N)$ operations for $N$ weakly-measured qubits. 
We demonstrate its efficiency by simulating the circuit dynamics of tens to hundreds of qubits in the area-law phase.

Another feature of our algorithm is that it returns the sampled measurements. This mimics what happens in real quantum hardware. As such, these sampled measurements can be safely utilized for prototyping purposes, 
before running experiments on real quantum hardware. Accessing these measurement samples may also be useful for validating quantum-computing hardware in the 
regime where efficient classical simulation is possible (e.g.~due to area-law entanglement). 

There are other settings where our algorithm may prove beneficial. Reproducing the samples of quantum measurement outcomes 
after a complex quantum evolution is known to be hard, and  this is the basis of some
quantum supremacy experiments \cite{arute2019quantum}.
More recently, real-world applications for these kinds of sampling problems have been 
studied for generative machine learning \cite{huang2025generative}. Here the samples 
represent the generated data, similar to large language models where text is what is generated \cite{radford2018improving}, or to image generation models \cite{ho2020denoising}. 
Tensor-network methods have already been applied to such generative-machine-learning tasks 
\cite{han2018unsupervised,stoudenmire2016supervised,wall2021generative,meiburg2025generative}. Our algorithm may find utility here, especially with its access to the sampled measurements.

The ability to perform weak measurements is also useful in applications such as 
temporal stochastic modelling. These are 
kinds of generative problems where the generated data is a time series, such as in stock market prices.  
Quantum simulators of temporal stochastic processes are attracting much attention 
since they can reproduce the same statistics of classical hidden Markov models, but with 
a much smaller memory
\cite{elliott2020extreme,wu2023implementing,glasser2019expressive,banchi2024accuracy,yang2025dimension}.
In their simplest form, these models can be expressed as a tensor network \textit{in time}. 
Our generalization to weak measurements is also an important step towards higher-dimensional problems, 
such as those involving both time and space. This is possible since the quantum memory is not destroyed and can be 
reused  to produce highly non-Markovian processes. 

The paper is structured as follow. In Sec.~\ref{sec:QCircuit} we define the setup of our quantum circuit. In Sec.~\ref{sec:MarkovChain} we explain our tensor-network 
algorithm and its Markov-chain structure. In Sec.~\ref{sec:Results} we test the performance of our algorithm 
by studying an instance of a MIPT, discussing its efficiency for area-law states. 
We present our conclusions and outlook in Sec.~\ref{sec:Conclusion}.

\section{Setup} 
\label{sec:Setup}

\subsection{The Quantum Circuit}
\label{sec:QCircuit}

\begin{figure}[t!]
  \centering    
  \begin{overpic}[width=\linewidth]{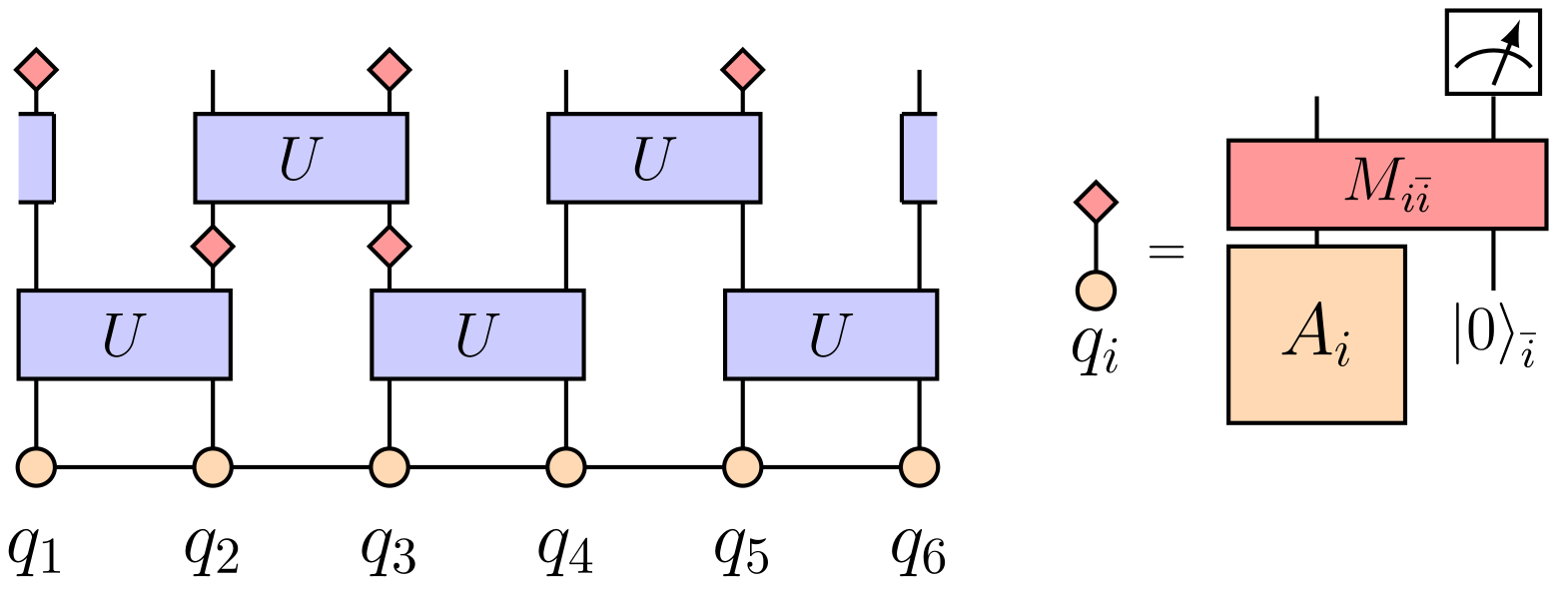}
    \put(-5,32){(a)}
    \put(65,32){(b)}
  \end{overpic}
  \caption{(a) The (1+1)-dimensional random unitary circuit studied in this work. The spatial direction consists of $N$ qubits (orange circles), with $N=6$ in this example. The temporal direction consists of Haar-random two-qubit unitaries $U$ in a brickwall structure (purple rectangles), interspersed with single-qubit weak measurements (red diamonds) with probability $p$ on a given qubit. The four layers shown here (two of random unitaries, two of weak measurements) constitute one time step. The unitaries $U$ are applied with periodic boundary conditions. (b) Explicit depiction of the weak measurement procedure of qubit $q_i$, where $M_{i\bar{i}}$ is given by Eq.~\eqref{eq:MeasGate}. Here, $\bar{i}$ is an ancilla qubit that is initialized in the $\ket{0}$ state and projectively measured after the application of $M_{i\bar{i}}$. $A_i$ is the tensor corresponding to the matrix-product state of qubit $q_i$, with link indices not shown.}
  \label{fig:QCircuit}
\end{figure}

\begin{figure*}[t!]
    \centering    
    \begin{overpic}[width=\linewidth]{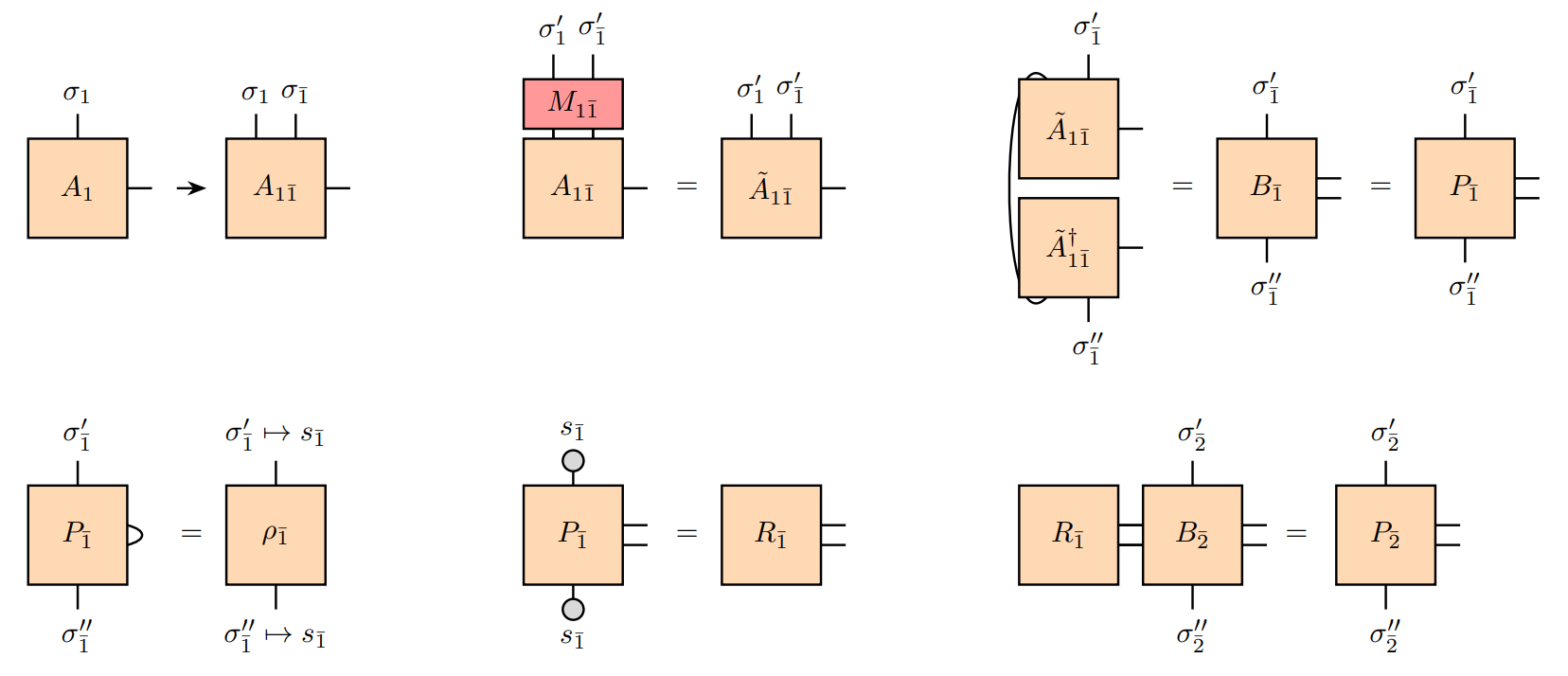}
      \put(0, 39){(a)}
      \put(30, 39){(b)}
      \put(62, 39){(c)}
      \put(0, 14){(d)}
      \put(30, 14){(e)}
      \put(62, 14){(f)}
    \end{overpic}
    \caption{The Markov chain approach to efficiently performing the weak measurements Eq.~\eqref{eq:MeasGate}. See the main text for full details. (a) Start with the first qubit, $q_1$. Form the tensor corresponding to the physical qubit's tensor $A_1$ and the ancilla qubit's initial state, $\ket{0}_{\bar{1}}$. Call this tensor $A_{1\bar{1}}$. (b) Apply the measurement operator Eq.~\eqref{eq:MeasGate}. Call the measurement-modified tensor $\tilde{A}_{1\bar{1}}$. (c) Contract the physical indices of the real qubit between $\tilde{A}_{1\bar{1}}$ and its Hermitian conjugate. Call this tensor $B_{\bar{1}}$ or $P_{\bar{1}}$. (d) Contract the right link indices of $P_{\bar{1}}$ and normalize it to form the one-particle density matrix for the ancilla qubit, $\rho_{\bar{1}}$. Sample a measurement outcome $s_{\bar{1}}$. (e) Return to the tensor $P_{\bar{1}}$. Take the component of this tensor corresponding to the measurement outcome $s_{\bar{1}}$, such that only right link indices remain. Call the resulting tensor $R_{\bar{1}}$. (f) For the next qubit, $q_2$, the procedure is similar. However, the tensor $B_{\bar{2}}$ will now possess left link indices. To form $P_{\bar{2}}$, contract $B_{\bar{2}}$ with $R_{\bar{1}}$ of the previous qubit. This propagates the previous measurement outcome forward in the Markov chain and along the matrix-product state.}
    \label{fig:MarkovMPS}
\end{figure*}
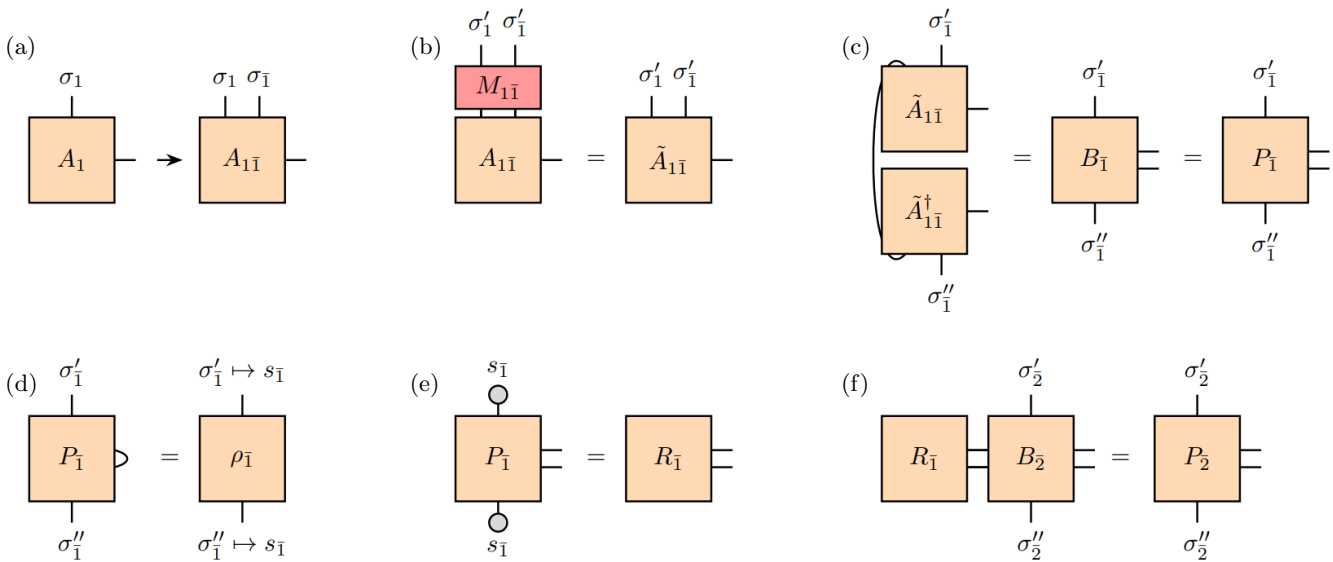

Although our algorithm can be applied to any quantum circuit, we focus here on the particular structure 
shown in Fig.~\ref{fig:QCircuit}. 
We consider a (1+1)-dimensional random unitary circuit as shown in Fig.~\ref{fig:QCircuit}(a). The spatial direction consists of $N$ qubits, initialized in a N\'eel state (i.e. alternating between $\ket{0}$ and $\ket{1}$) with periodic boundary conditions. The temporal direction is a brickwall structure of two-qubit random unitary operators $U$, interspersed with weak measurements on each qubit with probability $p$. We draw the random unitary operators $U$ from the Haar measure. Note that each unitary $U$, as shown in Fig.~\ref{fig:QCircuit}(a), will in principle be different.

The specific model we use for weak measurements of qubit $j$ is as follows. We first apply a unitary operation 
\begin{equation}
    M_{j\bar{j}} = \exp\left[i\theta \left( \frac{1+Z_j}{2} \right) X_{\bar{j}}\right], \label{eq:MeasGate}
\end{equation}
that couples qubit $j$ to an ancillary qubit $\bar j$. The ancillary qubits live outside of the chain of physical qubits $\{q_i\}$, shown in Fig.~\ref{fig:QCircuit}(a), and are initialized in the state $\ket{0}$. Performing a weak measurement means applying $M_{j\bar{j}}$ to a qubit and its ancilla, and then performing a projective measurement on the ancilla according to the Born rule; see Fig.~\ref{fig:QCircuit}(b). This approach to weak measurements is common in the quantum-information literature; see, for example, Ref.~\cite{BaoPRB2020}. When $\theta = 0$, the measurements are trivial and do not alter the wavefunction. When $\theta= \frac{\pi}{2}$, the measurements are projective. For $0 \leq \theta < \frac{\pi}{2}$, the measurements do not completely project the wavefunction and are therefore considered ``weak.'' Parameterizing measurements with the variable $\theta$ thus allows us to study a range of measurement strengths. Measurements of the form shown in Eq.~\eqref{eq:MeasGate} are implementable on current quantum computing architectures, such as the trapped-ion platform of Quantinuum; see Appendix \ref{sec:ZZGates}.  

\subsection{Markov-Chain Approach to Measurements}
\label{sec:MarkovChain}

We simulate the quantum circuit of Sec.~\ref{sec:QCircuit} using matrix-product state (MPS) techniques, as implemented in the ITensor library \cite{ITensor,ITensor-r0.3}. The construction of the initial state and the calculation of quantities of interest (i.e.~the bipartite entanglement entropy) are straightforward using these techniques. The sampling of Haar-random unitaries is directly implemented in ITensor as well, using the ``RandomUnitary'' designator in the \texttt{op()} function. Each layer of the brickwall structure of unitaries can be applied using the \texttt{apply()} function on the MPS. This function automatically performs the truncation of the MPS to the specified bond dimension and cutoff. For a review of MPS techniques, see Refs.~\cite{SchollwockRMP2005,SchollwockAnnPhys2011,CiracRMP2021}; some examples of using tensor networks to simulate quantum circuits can be found in Refs.~\cite{MelkoPRA2021, HaghshenasPRX2022,SeitzQuantum2023, BerezutskiiArxiv2025}. 

In this section, we demonstrate how to efficiently perform weak measurements of the quantum circuit using a Markov chain in tandem with MPS contractions, which is the main focus of this work. The form of weak measurements Eq.~\eqref{eq:MeasGate} can be efficiently evaluated using a Markov chain. The procedure is depicted in Fig.~\ref{fig:MarkovMPS} and is as follows:
\begin{enumerate}
    \item Begin with the first qubit, $i=1$, with tensor $A_1$. Set this as the orthogonality centre. Construct the corresponding tensor with the ancilla included, $A_{1\bar{1}}$. See Fig.~\ref{fig:MarkovMPS}(a).
    \item Apply the measurement operator $M_{1\bar{1}}$ to this tensor. Call the resulting tensor $\tilde{A}_{1 \bar{1}}$. See Fig.~\ref{fig:MarkovMPS}(b).\label{item:Meas}
    \item  Contract the real qubit's physical index in $\tilde{A}_{1 \bar{1}}$ with that of its Hermitian conjugate. Call the resulting tensor $B_{\bar{1}}$, or -- as we will clarify shortly -- $P_{\bar{1}}$. See Fig.~\ref{fig:MarkovMPS}(c).  \label{item:Next}
    \item Form the one-particle density matrix for the ancilla qubit by contracting the dangling right link indices of $P_{\bar{1}}$ and normalizing. Call the resulting tensor $\rho_{\bar{1}}$. Sample a measurement outcome $s_{\bar{1}}$ from this density matrix. See Fig.~\ref{fig:MarkovMPS}(d).
    \item Project both the ancilla indices of $P_{\bar{1}}$ into the measured outcome $s_{\bar{1}}$. The resulting tensor only has right link indices. Call this tensor $R_{\bar{1}}$. See Fig.~\ref{fig:MarkovMPS}(e).
\end{enumerate}
This procedure repeats almost identically for the next qubit, $i=2$. However, in step \ref{item:Next}, the tensor $B_{\bar{2}}$ that is formed has left link indices as well as right link indices. At this step, $B_{\bar{2}}$ is contracted from the left with $R_{\bar{1}}$, the tensor that captures the measurement outcome on the previous qubit. This is what defines $P_{\bar{2}}$. See Fig.~\ref{fig:MarkovMPS}(f). 

At the end of this Markov chain, one has a collection of sampled measurement outcomes $\{s_{\bar{i}}\}$ for all ancilla qubits. One then projects the ancilla indices of the MPS onto these sampled outcomes, representing projective measurements of the ancilla qubits. The resulting MPS is the state of the quantum circuit after the layer of measurements. If some qubits are to be left unmeasured, one simply omits projecting onto the samples for those qubits. Equivalently, one could follow the procedure outlined in Fig.~\ref{fig:MarkovMPS} for such qubits, but take $M_{i\bar{i}}$ to be the identity.

The tensor $P_{\bar{i}}$ is responsible for propagating the result of measurements forward to the next tensor and captures the essence of the Markov-chain approach. Sampling the measurement outcomes one tensor at a time reduces the complexity of subsequent MPS contractions, which is the main computational benefit.

\section{Results}
\label{sec:Results}

\begin{figure*}[t!]
    \centering 
    \begin{overpic}[width=\linewidth]{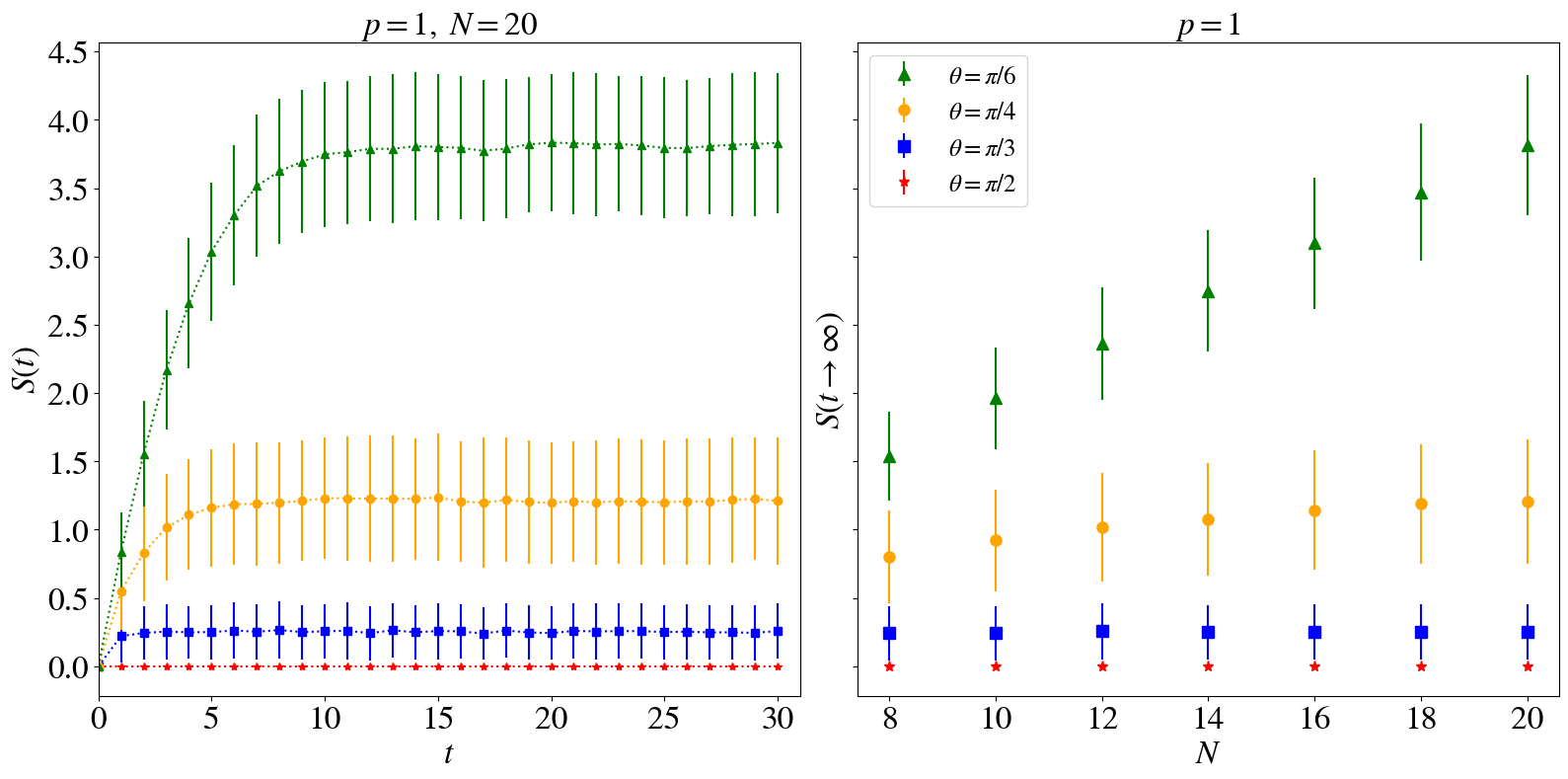}
      \put(0,48){(a)}    
      \put(51,48){(b)}    
    \end{overpic}   
    \caption{(a) Time series of $S(t)$ within the quantum circuit, for $p=1$, $N=20$, and various measurement strengths: $\theta=\pi/2$ (red downwards triangles), $\theta=\pi/3$ (blue squares), $\theta=\pi/4$ (orange circles), and $\theta=\pi/6$ (green upwards triangles). The averaging procedure over $N_{\rm} = 1000$ quantum trajectories is described in the main text. Error bars represent one standard deviation. (b) Long-time entanglement entropy $S(t\to\infty)$ for the same measurement rate $p$ and measurement strengths $\theta$ as in (a), but as a function of system size $N$. Error bars correspond to the standard error of the mean.}
    \label{fig:Transition}
\end{figure*}

In this section, we present demonstrative results of our MPS algorithm for various choices of qubit number $N$ and measurement strength $\theta$, highlighting the phenomenon of MIPTs as an application. Unless otherwise stated, we restrict ourselves to the case of weakly measuring all qubits (i.e.~$p=1$), set the maximum bond dimension in the simulations to be $\chi = 1000$, and set the cutoff for singular-value-decomposition truncations in ITensor as $\epsilon = 10^{-6}$. All simulations are run on an Intel Xeon Gold 6526Y processor using 32 threads.

We are interested in studying the bipartite entanglement entropy,
\begin{equation}
    S = -\Tr\left[\rho_{\rm A} \ln(\rho_{\rm A})\right], \label{eq:BPEE}
\end{equation}
for some subsystem A with reduced density matrix $\rho_{\rm A}$. We are specificially interested in calculating $S$ as a function of time $t$, as well as the number of qubits $N$ and the measurement strength $\theta$. Here, one unit of time consists of two layers of random unitaries and two rounds of measurements, as depicted in Fig.~\ref{fig:QCircuit}(a). To evaluate Eq.~\eqref{eq:BPEE}, we take $N$ to be even and divide the chain into two subsystems. As noted in a similar setting in Ref.~\cite{SzyniszewskiPRB2019}, an equal cut performed at site $\frac{N}{2}$ leads to differences between chains of length $4n$ and $4n+2$ for natural numbers $n$; this is because such cuts may or may not be commensurate with a given layer of random unitaries. To treat all chain lengths $N$ on equal footing, we therefore calculate Eq.~\eqref{eq:BPEE} with a cut at $\frac{N}{2}$ as well as with a cut at $\frac{N}{2}+1$, and average the two results. We do this at every time $t$ for $N_{\rm runs} = 1000$ distinct quantum trajectories, and average over all trajectories to produce the final estimation of the entanglement entropy at time $t$, $S(t)$. In order to estimate the long-time limit of the entanglement entropy, we average $S(t)$ over all times $t \geq t_{\rm cutoff}$; we choose $t_{\rm cutoff} = 20$, and we call this long-time limit $S(t \to \infty)$. 

As a preliminary check, we first take $\theta=\pi/2$ and $p=0.33$. $\theta=\pi/2$ corresponds to projective measurements, and this choice of $p$ is within the area-law phase \cite{skinner2019measurement,MelkoPRA2021}. This limit is therefore easy to simulate directly with tensor networks using the Born rule for measurements. We calculate $S(t \to \infty)$ as a function of $N$, from $N=8$ to $N=20$, using this direct approach as well as the Markov-chain approach outlined in Fig.~\ref{fig:MarkovMPS}. We find excellent agreement between the two.

We next apply our Markov-chain approach to a variety of system sizes $N$ and measurement strengths $\theta$, choosing $p=1$. We specifically simulate the quantum circuit for $N = 8, 10, \dots, 20$ and $\theta/\pi = 2^{-1}, 3^{-1}, 4^{-1}, 6^{-1}$, in the order of strongest to weakest measurement strengths. The results are shown in Fig.~\ref{fig:Transition}. Figure~\ref{fig:Transition}(a) shows the time series of the entanglement entropy $S(t)$ for the largest system size, $N=20$, and all measurement strengths. Figure~\ref{fig:Transition}(b) shows the long-time entanglement entropy $S(t\to\infty)$ for all system sizes $N$ and measurement strengths. In contrast to Ref.~\cite{MelkoPRA2021}, the error bars in Fig.~\ref{fig:Transition}(a) correspond to the standard deviation rather than the standard error of the mean. We do this to quantify the fluctuations of individual runs, rather than their average. On the other hand, the error bars in Fig.~\ref{fig:Transition}(b) correspond to the standard error of the mean.

\begin{figure}[t!]
  \centering    
  \includegraphics[width=\linewidth]{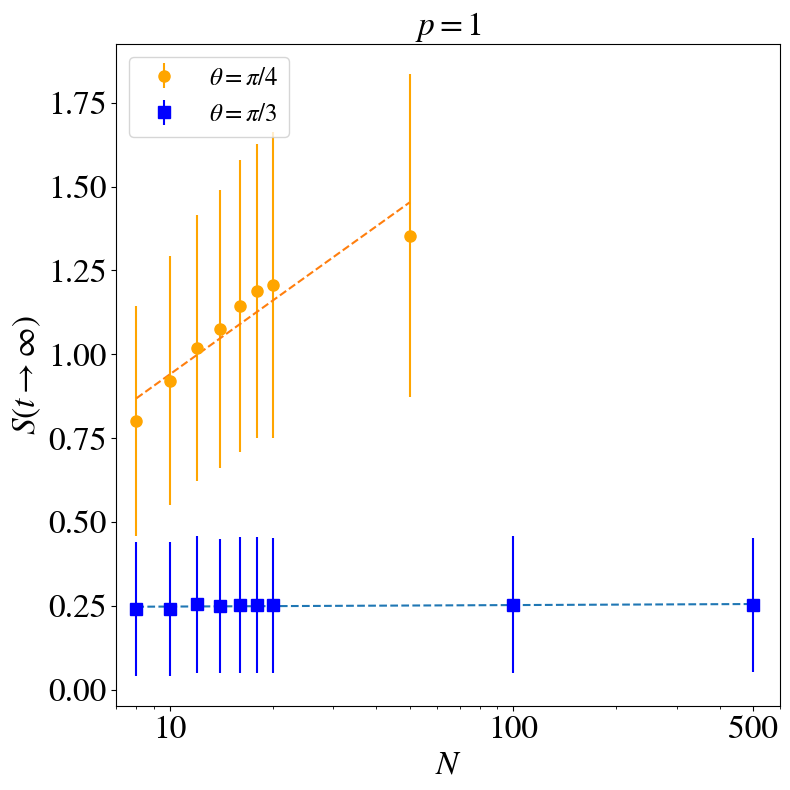}
  \caption{Long-time entanglement entropy $S(t\to\infty)$ for $p=1$ and measurement strengths $\theta = \pi/4, \pi/3$. The data is the same as in Fig.~\ref{fig:Transition}(b), with the inclusion of larger system sizes: $N = 50$ ($\theta = \pi/4$) and $N=100,\ 500$ ($\theta=\pi/3$). Error bars correspond to the standard error of the mean. The dashed lines show lines of best fit for each choice of $\theta$, corresponding to a $\log(N)$ scaling of the long-time entanglement entropy.}
  \label{fig:NScaling}
\end{figure}

As expected for $\theta=\pi/2$ (projective measurements), the time series as well as the long-time entanglement entropy are always zero regardless of system size, since every qubit is measured. As the measurement strength decreases, $S(t \to \infty)$ takes on a finite value. For $\theta = \pi/3$, measurements are still sufficiently strong and frequent to make $S(t\to\infty)$ virutally independent of system size $N$. For $\theta=\pi/4$, there is a weak dependence of $S(t\to\infty)$ on the system size. We fit the data to a logarithmic function in $N$, finding excellent agreement (not pictured). On the other hand, $\theta=\pi/6$ shows a clear linear dependence on system size. We therefore see a transition from area-law entanglement ($S \sim \log(N)$) to volume-law entanglement ($S \sim N$) purely caused by a change in measurement strength, as has been studied before \cite{SzyniszewskiPRB2019}. This is in contrast to the original work on MIPTs \cite{skinner2019measurement}, which found a transition from area- to volume-law entanglement by changing the \textit{rate} of measurements. In our setting, the transition can be captured with tensor-network simulations on moderate system sizes using modest computational resources. We leave a detailed exploration of this phase transition for future work.

To demonstrate the efficiency of our algorithm, we repeat the same simulations for measurement strengths $\theta = \pi/4$ and $\theta=\pi/3$, but this time for a much greater number of qubits. Our results are displayed in Fig.~\ref{fig:NScaling}; we plot the same data as Fig.~\ref{fig:Transition}(b) for $\theta = \pi/4$ and $\theta=\pi/3$, as well as new data for $N = 50$ ($\theta = \pi/4$) and $N=100,\ 500$ ($\theta=\pi/3$). We present the results on a logarithmic scale in $N$, such that the expected area-law behavior should present itself as a linear relation. The line of best fit through each dataset is consistent with this expectation, within error bars. These simulations again only involve a moderate amount of computational resources: when run on the aforementioned processor, all simulations in Fig.~\ref{fig:Transition} and Fig.~\ref{fig:NScaling} took as little as a few hours to as much as a few days. The bottleneck in these simulations is the level of entanglement and not merely the system size. For example, the $N=500$ simulation for $\theta=\pi/3$ required only a few hours, whereas the the $N=50$ simulation for $\theta= \pi/4$, and the $N=20$ simulation for $\theta=\pi/6$, required a few days. This bottleneck is understandable for a tensor-network-based method, which becomes more computationally expensive with increased entanglement and not necessarily so with increased system size.

\begin{figure}[t!]
  \centering    
  \includegraphics[width=\linewidth]{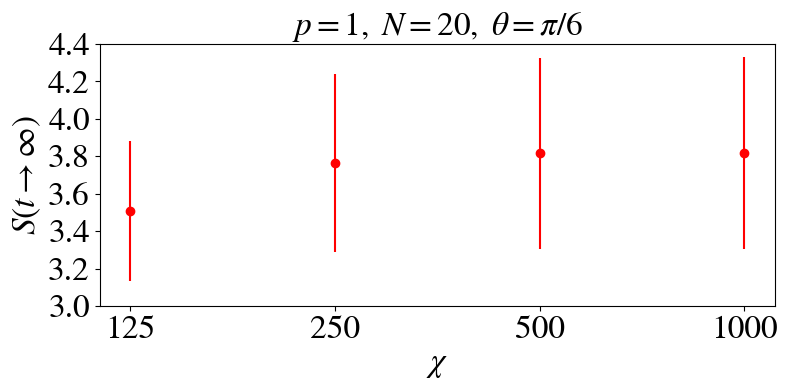}
  \caption{Long-time entanglement entropy $S(t\to\infty)$ for $p=1$, $N=20$, and $\theta=\pi/6$, as a function of bond dimension $\chi$ in the MPS simulations. Error bars correspond to the standard error of the mean.}
  \label{fig:BDScaling}
\end{figure}

Finally, we monitor the scaling of $S(t \to \infty)$ with bond dimension $\chi$ at $p = 1$, $N=20$, and $\theta=\pi/6$; see Fig.~\ref{fig:BDScaling}. These parameters correspond to the volume-law entanglement scaling shown in Fig.~\ref{fig:Transition}(b). There is a convergence of $S(t\to\infty)$ as the bond dimension increases, showing little difference between $\chi=500$ and $\chi=1000$. Our results for the entanglement entropy in Fig.~\ref{fig:Transition} can therefore be trusted, even with the singular value decompositions and MPS truncations that are necessary at $\chi=1000$. 

\section{Summary \& Outlook}
\label{sec:Conclusion}

We have developed a tensor-network algorithm for simulating the dynamics of monitored quantum circuits, with mid-circuit measurements whose strength is parameterized by an angle $\theta$ (Eq.~\eqref{eq:MeasGate}). The crux of the algorithm is to use a Markov chain to sample measurement outcomes and feed their effect forward along the chain of qubits, allowing the tensor network to be more efficiently contracted. As a demonstration, we simulated the dynamics for $\theta = \pi/2, \pi/3$, and $\pi/4$, which exhibit area-law entanglement, on tens to hundreds of qubits. We also simulated the dynamics for $\theta=\pi/6$ for tens of qubits, which exhibit volume-law entanglement and therefore hint at a MIPT as a function of $\theta$. Previous works \cite{SzyniszewskiPRB2019,MelkoPRA2021} have shown that tens of qubits is a sufficient size for analyzing MIPTs. Features of this transition should therefore be accessible to tensor-network simulations using our algorithm and requiring only a modest amount of computational time and resources. This is a straightforward direction for future work.

Although the phenomenon of MIPTs was only presented as a particular use case of our algorithm, it is worth mentioning its realizations in real quantum hardware. The stochastic nature of quantum measurement outcomes poses a formidable challenge known as the ``post-selection problem'' \cite{McGinleyPRX2024,GarrattPRX2024}, though several workarounds have been proposed 
\cite{BawejaArxiv2024,FengArxiv2025, BuchholdArxiv2022,KhindanovAnnPhys2025,GarrattPRX2024}. As well, although tensor-network methods are generally limited to quantum systems with low levels of entanglement (e.g. circuit dynamics with area-law scaling), there are ways to circumvent or even utilize this limitation. For example, Ref.~\cite{CecilePhysRevRes2024} showed that the error between true and MPS-approximated time evolution has different scaling behavior (with the bond dimension $\chi$) for volume- and area-law phases. Hence, scaling arguments from small bond dimensions can provide insight into the MIPT. Similarly, Ref.~\cite{YanayPRL2024} proposed a hybrid quantum-classical algorithm which uses the \textit{failure} of MPS representations in the volume-law phase to pinpoint the measurement-induced critical point. It is conceivable that our tensor-network algorithm may be integrated with such workarounds, especially given its mimicry of real quantum hardware and its access to mid-circuit outcomes of variable-strength measurements. 

Taking a step back, our algorithm is general. We have profiled it using the phenomenon of measurement-induced criticality; however, it can  be used to study the dynamics of monitored quantum circuits in other settings, with applications in generative machine learning or the modelling of complex temporal processes. The only feature that is needed is the Markov-chain structure, to sample measurements and to simplify the tensor-network contractions. This could reasonably be present in other measurement protocols. As a tensor-network representation of the quantum circuit, our algorithm is also open to that whole world of techniques. This includes, for example, making use of tensor-network methods for calculating quantities like fidelities \cite{HauruPRA2018}. 

Tensor networks are naturally powerful tools for effectively representing high-dimensional data. However, algorithmic developments (rather than hardware improvements) are often what push the boundary of what they can simulate. This is highlighted, for example, by the use of belief propagation for studying the disordered dynamics of quantum glasses \cite{TindallArxiv2025} or clever contraction and sampling methods for avoiding large entanglement barriers in quantum dynamics \cite{CarignanoArxiv2025}. Our algorithm is one more contribution to this space, and should provide a fresh avenue for simulating the dynamics of quantum circuits with weak mid-circuit measurements.

\section*{acknowledgements}
D.P.~acknowledges the support of the Natural Sciences and Engineering Research Council of Canada (NSERC) (Ref. No. PGSD-567963-2022).
L.B.~acknowledges financial support from: 
 PNRR Ministero Universit\`a e Ricerca Project No. PE0000023-NQSTI, funded by European Union-Next-Generation EU; 
 Prin 2022 - DD N. 104 del 2/2/2022, entitled ``understanding the LEarning process of QUantum Neural networks (LeQun)'', proposal code 2022WHZ5XH, CUP B53D23009530006; 
 the European Union's Horizon Europe research and innovation program under EPIQUE Project GA No. 101135288. 


\appendix

~

\section{Weak Measurements from Native Gate Set}
\label{sec:ZZGates}

In this Appendix, we describe how to implement the gate Eq.~\eqref{eq:MeasGate} using the native gates of trapped-ion platforms. We take Quantinuum's trapped-ion quantum computer, System Model H2, as our typical example \footnote{\label{eq:Quantinuum}For more information on Quantinuum's trapped-ion platform, System Model H2, see their product data sheet \href{https://www.quantinuum.com/products-solutions/quantinuum-systems/system-model-h2}{here}.}. In its native gate set are single-qubit rotations, such as $e^{i\theta Z_i}$ on qubit $i$, as well as two-qubit $ZZ$ gates, $e^{i \theta Z_i Z_j}$. Hence we can combine these to produce
\begin{equation}
  \mathcal{M}_{ij}(\theta) = \exp[i\theta \frac{Z_i}{2}] \exp[i\theta \frac{Z_i Z_j}{2}].
\end{equation}
By performing a single-qubit rotation on qubit $i$ (i.e. using a single-qubit rotation with a Hadamard gate), this can be mapped onto
\begin{align}
  \mathcal{M}_{ij}(\theta) &\to \exp[i\theta \frac{X_i}{2}] \exp[i\theta \frac{X_i Z_j}{2}] \\
  &= \exp[i\theta \left(\frac{1 + Z_j}{2}\right)X_i].
\end{align}
By relabelling $i \to \bar{j}$ for the ancilla qubit, we recover Eq.~\eqref{eq:MeasGate}. For more information on the gates available to trapped-ion platforms, particularly those relevant to MIPTs, see Ref.~\cite{MelkoPRA2021}.

\bibliography{refs.bib}
\end{document}